\documentclass[final,1p,times]{elsarticle}

\usepackage{siunitx}
\usepackage{amssymb}
\usepackage{svg}
\usepackage{threeparttable}
\usepackage{booktabs}
\usepackage{caption}
\usepackage{amsmath}
\usepackage{hyperref}
\usepackage{graphicx}
\usepackage{subfigure}
\usepackage{algorithm}
\usepackage{algpseudocode}
\usepackage{longtable}
\usepackage{mathrsfs}
\usepackage{lineno}
\usepackage{multirow}
\usepackage{amsthm}
\newtheorem{lemma}{Lemma}

\journal{Journal of Computational Science}

\begin{document}

\begin{frontmatter}



\title{HybridOctree\_Hex: Hybrid Octree-Based Adaptive All-Hexahedral Mesh Generation with Jacobian Control}


\author[label1]{Hua Tong}
\author[label1,label2,label3]{Eni Halilaj}
\author[label1,label2]{Yongjie Jessica Zhang}

\address[label1]{Department of Mechanical Engineering, Carnegie Mellon University, 5000 Forbes Ave, Pittsburgh, PA 15213, USA}
\address[label2]{Department of Biomedical Engineering, Carnegie Mellon University, 5000 Forbes Ave, Pittsburgh, PA 15213, USA}
\address[label3]{Robotics Institute, Carnegie Mellon University, 5000 Forbes Ave, Pittsburgh, PA 15213, USA}


\begin{abstract}

We present a new software package, ``HybridOctree\_Hex,'' for adaptive all-hexahedral mesh generation based on hybrid octree and quality improvement with Jacobian control. The proposed HybridOctree\_Hex begins by detecting curvatures and narrow regions of the input boundary to identify key surface features and initialize an octree structure. Subsequently, a strongly balanced octree is constructed using the balancing and pairing rules. Inspired by our earlier preliminary hybrid octree-based work, templates are designed to guarantee an all-hexahedral dual mesh generation directly from the strongly balanced octree. With these pre-defined templates, the sophisticated hybrid octree construction step is skipped to achieve an efficient implementation. After that, elements outside and around the boundary are removed to create a core mesh. The boundary points of the core mesh are connected to their corresponding closest points on the surface to fill the buffer zone and build the final mesh. Coupled with smart Laplacian smoothing, HybridOctree\_Hex takes advantage of a delicate optimization-based quality improvement method considering geometric fitting, Jacobian and scaled Jacobian, to achieve a minimum scaled Jacobian that is higher than $0.5$. 
We empirically verify the robustness and efficiency of our method by running the HybridOctree\_Hex software on dozens of complex 3D models without any manual intervention or parameter adjustment. We provide the HybridOctree\_Hex source code, along with comprehensive results encompassing the input and output files and statistical data in the following repository: \url{https://github.com/CMU-CBML/HybridOctree_Hex}.

\end{abstract}



\begin{keyword}
All-hexahedral mesh generation\sep hybrid octree\sep dual mesh\sep adaptive mesh\sep Jacobian\sep quality improvement



\end{keyword}

\end{frontmatter}

\section{Introduction}

Hexahedral (hex) mesh generation is a widely adopted volumetric discretization method that is crucial for solving partial differential equations in diverse fields such as computer graphics, medical modeling, and engineering simulations \cite{zhang2016geometric}. Compared with tetrahedral meshes, all-hex elements are often preferred due to their superior performance in terms of increased accuracy, smaller element counts, and improved reliability \cite{benzley1995comparison, shepherd2006quality}. Despite the acknowledged benefits of hex meshes, the automatic generation of high-quality, conforming meshes remains a challenging problem \cite{owen1998survey, zhang2013challenges}. Due to the geometrical stiffness of hex elements, it is difficult to introduce local modifications or adapt mesh refinement strategies compared to quadrilateral (quad) or tetrahedral meshes \cite{schneiders2000algorithms}. Existing techniques for automation have significant trade-offs, and there is a need for further research to overcome technical hurdles and develop improved hex-meshing algorithms and software packages \cite{blacker2000meeting}. The complexities also involve decomposing geometry, handling assembly models with multiple components and materials, and propagating the mesh across shared surfaces. Addressing these challenges requires sophisticated algorithms and tools \cite{tautges2001generation}. In addition, all-hex mesh generation poses significant challenges due to the complexity of satisfying geometric and topological constraints, the limitations of current algorithms in handling diverse and complex geometries, and the open questions regarding the untangling of elements, prediction of element quality, and extension of meshing algorithms to implicitly satisfy a broader range of constraints \cite{shepherd2008hexahedral}. In addition, all-hex meshing also requres sharp feature preservation, robust quality improvement, and high-order element construction for intricate domains \cite{shepherd2007quality, zhang2013challenges}. 

Recent research efforts have largely focused on developing polycube or cross-field methods. The volumetric polycube method facilitates the conversion of a volume to an all-hex mesh by initially embedding axis-aligned boxes within the 3D volume, followed by mapping the volume onto multiple interconnected boxes \cite{gregson2011all}. The modified centroidal Voronoi tessellation is taken into account in the space of normals or eigenfunctions to segment the surface and construct polycubes \cite{hu2016centroidal, hu2017surface}. The input model can also be parameterized into a polycube structure to integrate geometry design with isogeometric analysis \cite{yu2022hexgen}. CubeCover employs a valid 3D frame field to globally parametrize and generate all-hex meshes \cite{nieser2011cubecover}. A sophisticated all-hex meshing framework is introduced, which is guided by a singularity-restricted field for volume parameterization \cite{li2012all}. The octahedral field method is improved with singularity graph correction and hex-meshable constrained octahedral field \cite{liu2018singularity}. While these methods are capable of producing near-optimal meshes that follow sharp features, they often struggle to guarantee all-hex or adaptive meshes or to avoid inverted elements.

In recent years, the generation of all-hex meshes using octree data structures is the only solution with the capability to satisfy rigorous scalability and robustness criteria for arbitrary shapes based on adaptive Cartesian grids. Extending the dual contouring method to achieve crack-free interval volume 3D meshing with boundary feature-sensitive adaptation enables the rapid extraction of adaptive and high-quality 3D finite element meshes from volumetric imaging data \cite{zhang20053d, zhang2006adaptive}. Octree data structure is exploited to generate high-quality meshes for implicit solvation models of biomolecular structures \cite{zhang2006quality}. Grid-based methods have proven effective in addressing multi-material scenarios \cite{zhang2008automatic, zhang2010automatic}, and they are capable of managing intricate non-manifold computer-aided design (CAD) assemblies as well \cite{qian2012automatic}. An integrated tool, LBIE-Mesher, is designed for constructing 2D or 3D finite element meshes from improved images, supporting features such as multiscale modeling, automatic mesh generation for heterogeneous domains, sharp feature preservation in all-hex meshes, robust quality improvement for non-manifold meshes, and construction of high-order elements \cite{zhang2016geometric, zhang2013challenges}.

In the octree-based pipeline, we need to fit a 3D grid of hexes in the volume and then add hex elements at the boundaries to fill gaps or create a dual mesh. The interior hexes are filled with templates (dense at the boundary and coarse inside), whereas the boundary element quality has no guarantee. There are two pervasive challenges in octree-based methods. The first challenge is the occurrence of hanging nodes, which arise when neighboring cells possess differing resolution levels, leading to nodes positioned in the middle of adjacent faces or edges. An algorithm that resolves hanging nodes by connecting them with polyhedra has been previously introduced, enabling the extraction of all-hex meshes as duals of polyhedral cells \cite{marechal2009advances}. The cutting process is improved to reduce the number of elements in the interior mesh \cite{hu2013adaptive}. The same dual mesh scheme is adopted in \cite{gao2019feature}, but with a different approach, whereas the templates are not thoroughly covered in the paper. By weakly balancing the octree, the class of adaptive meshes can be expanded, thereby relaxing the strong topological constraints. This leads to an enhanced convergence towards valid solutions, optimization of the number and distribution of singular mesh edges, and a reduction in the element count \cite{livesu2021optimal}. The generalized pairing criterion is capable of notably decreasing both grid and mesh size while ensuring to produce an all-hex mesh \cite{pitzalis2021generalized}. 

The second challenge is to maintain the inversion-free property of the surface mesh. Laplacian smoothing is a straightforward and easy-to-implement technique that repositions a vertex to the average of its neighbors \cite{canann1998approach}. While this method is cost-effective and simple to implement, it carries the risk of inverting surrounding elements. To address this challenge, an optimization-based approach that assesses the quality of elements surrounding a node and strives to improve them is proposed \cite{zhang2006adaptive, zhang2009surface}. A hybrid approach, integrating Laplacian smoothing with optimization, is advocated to balance efficiency and robustness \cite{freitag1997combining, canann1998approach}. A novel method for enhancing the quality of non-manifold hex meshes in microstructure materials, employing a vertex categorization approach and a comprehensive technique involving pillowing, geometric flow, and optimization, addresses challenges in previous works \cite{qian2010quality}. Iterative smoothing techniques are proposed to gradually relocate vertices towards the boundary, prohibiting local smoothing adjustments in case of hex flipping \cite{marechal2009advances, lin2015quality}. These methods are computationally efficient but may occasionally fail to preserve the geometry. A global deformation method exhibits robust alignment of the resulting mesh with the input surface with sharp features and ensures the alignment within an error bound \cite{gao2019feature}. However, the robustness comes at the cost of increased computational and memory resources, which might be prohibitive for standard hardware configurations.

In this paper, we introduce a new software package, ``HybridOctree\_Hex,'' designed for the automatic, robust, and efficient generation of adaptive and quality all-hex meshes with Jacobian control. Our method ensures the absence of self-intersections and produces high scaled Jacobians ($> 0.5$), which well surpass the minimum scaled Jacobian of interior template elements. Moreover, this innovation is particularly adept at capturing intricate, detailed features with precision. Given a 3D closed manifold surface as the starting point, our method first initializes an octree structure and builds a strongly balanced octree. Inspired by our earlier preliminary hybrid octree mesh generation technique \cite{hu2013adaptive}, we catalog all possible transition scenarios encountered during the all-hex dual mesh generation process into pre-defined templates. This comprehensive approach allows better mesh adaptation and exhibits faster template transformation. Subsequently, we remove elements outside and around the boundary and then 
connect the boundary points of the interior core mesh with their corresponding closest points on the boundary surface to fill the buffer zone. During the final step of meshing the buffer zone, we incorporate geometric fitting, the scaled Jacobian, and the Jacobian into our energy function, leverage the smart Laplacian smoothing to expedite convergence, and address issues of surface points getting trapped in local minima. Our extensive experimental results, obtained by processing dozens of models, serve as a testament to the robustness and efficacy of our proposed HybridOctree\_Hex package. In comparison to the current state-of-the-art technique outlined in \cite{gao2019feature}, our method produces superior-quality meshes with a reduced number of elements. To facilitate further research and collaboration, we have made available our HybridOctree\_Hex software source code and a comprehensive collection of meshes generated using our algorithm, along with corresponding input and output data, at \url{https://github.com/CMU-CBML/HybridOctree_Hex}.

The paper structure is outlined as follows: Section \ref{sec:2} delineates the comprehensive algorithm for initializing an octree, generating a strongly balanced octree, designing templates based on our earlier preliminary hybrid octree work, generating the interior core mesh, and meshing the buffer zone with geometric fitting Jacobian control. Section \ref{sec:3} presents meshing results and Section \ref{sec:4} concludes the paper and outlines future research.

\section{Hybrid Octree and All-Hex Mesh Generation}
\label{sec:2}

\begin{figure}
  \centering
  \subfigure[]{
    \includegraphics[width=0.33\textwidth]{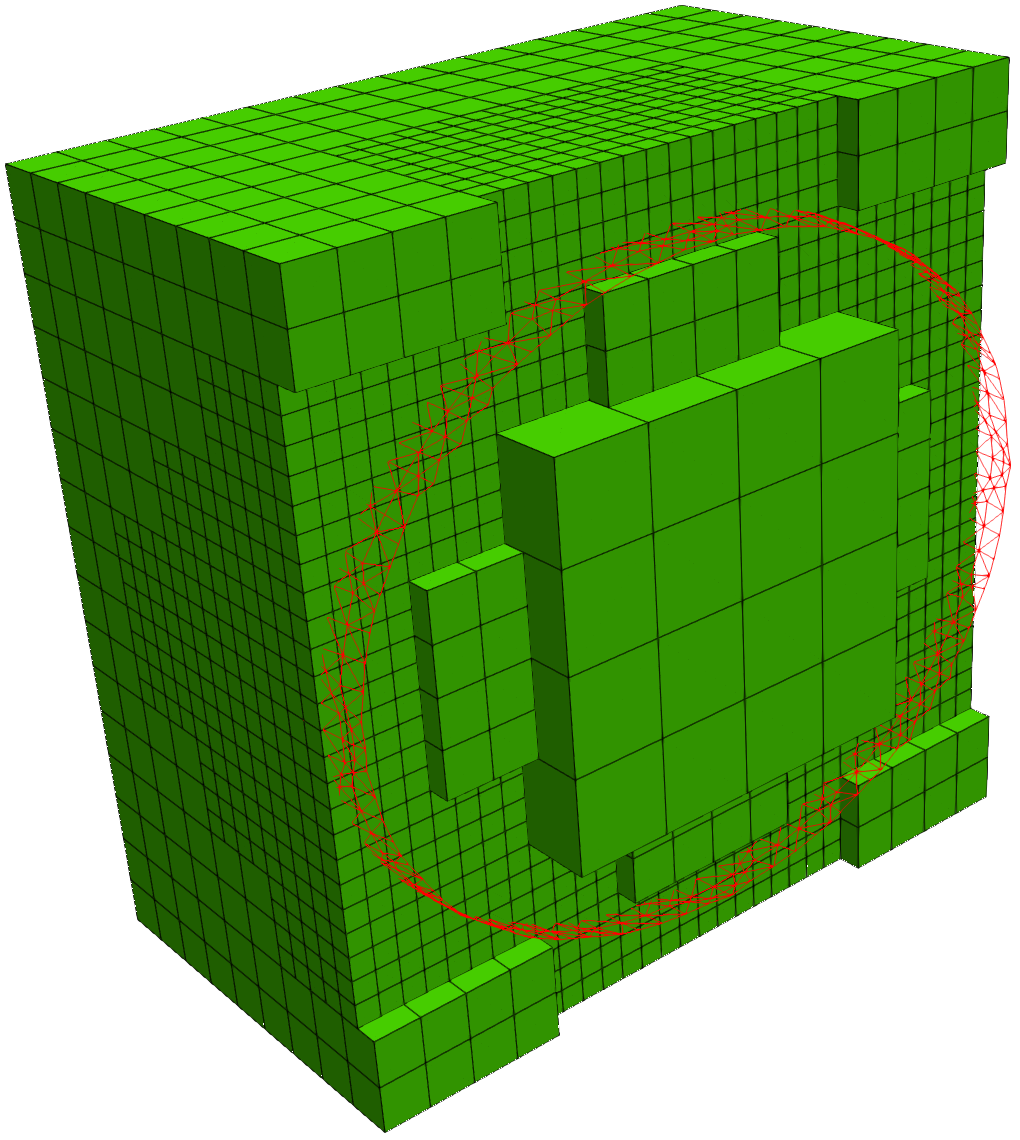}
    \label{fig:octree0}  
  }\hspace{-4mm}
  \subfigure[]{
    \includegraphics[width=0.33\textwidth]{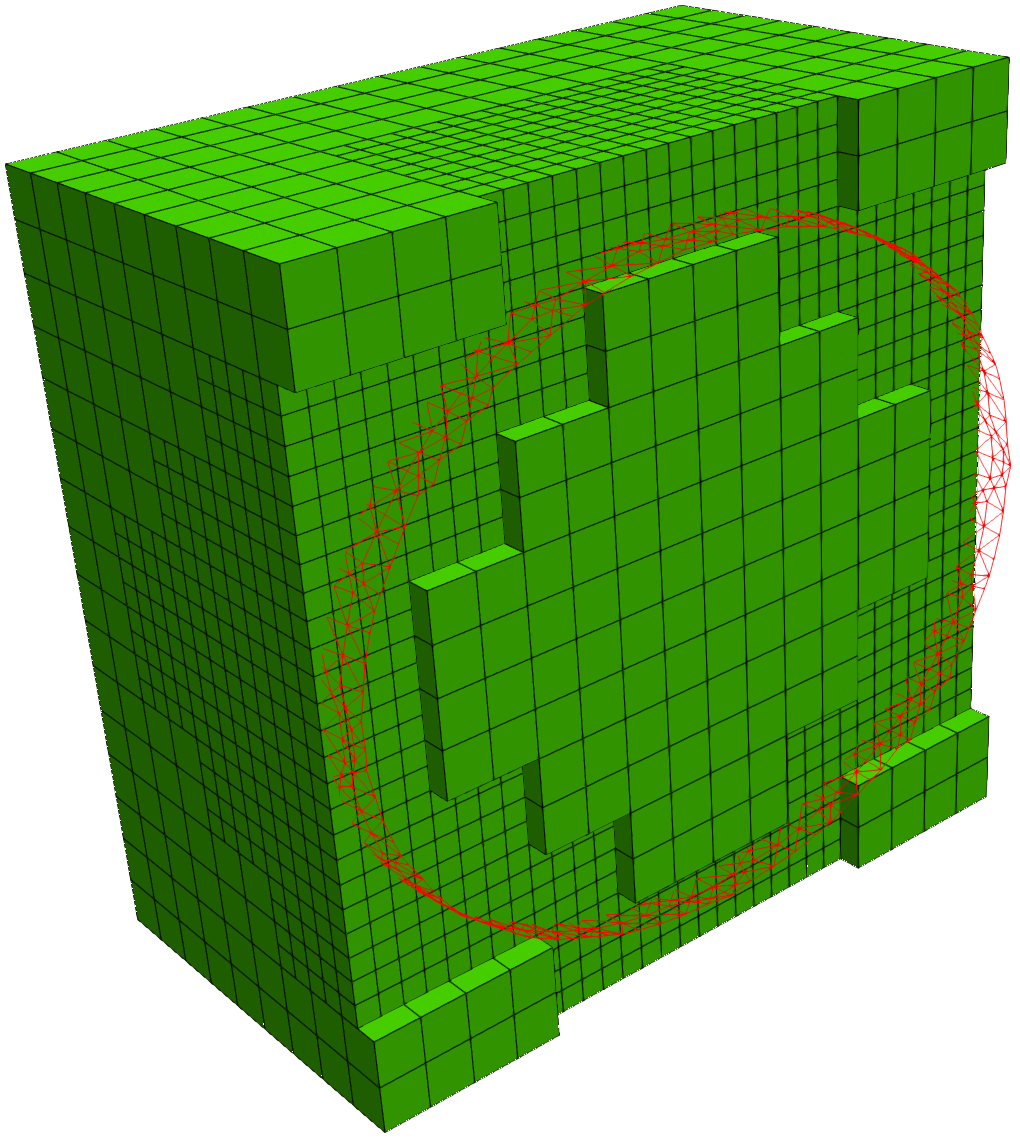}
    \label{fig:octree}  
  }\hspace{-4mm}
  \subfigure[]{
    \includegraphics[width=0.33\textwidth]{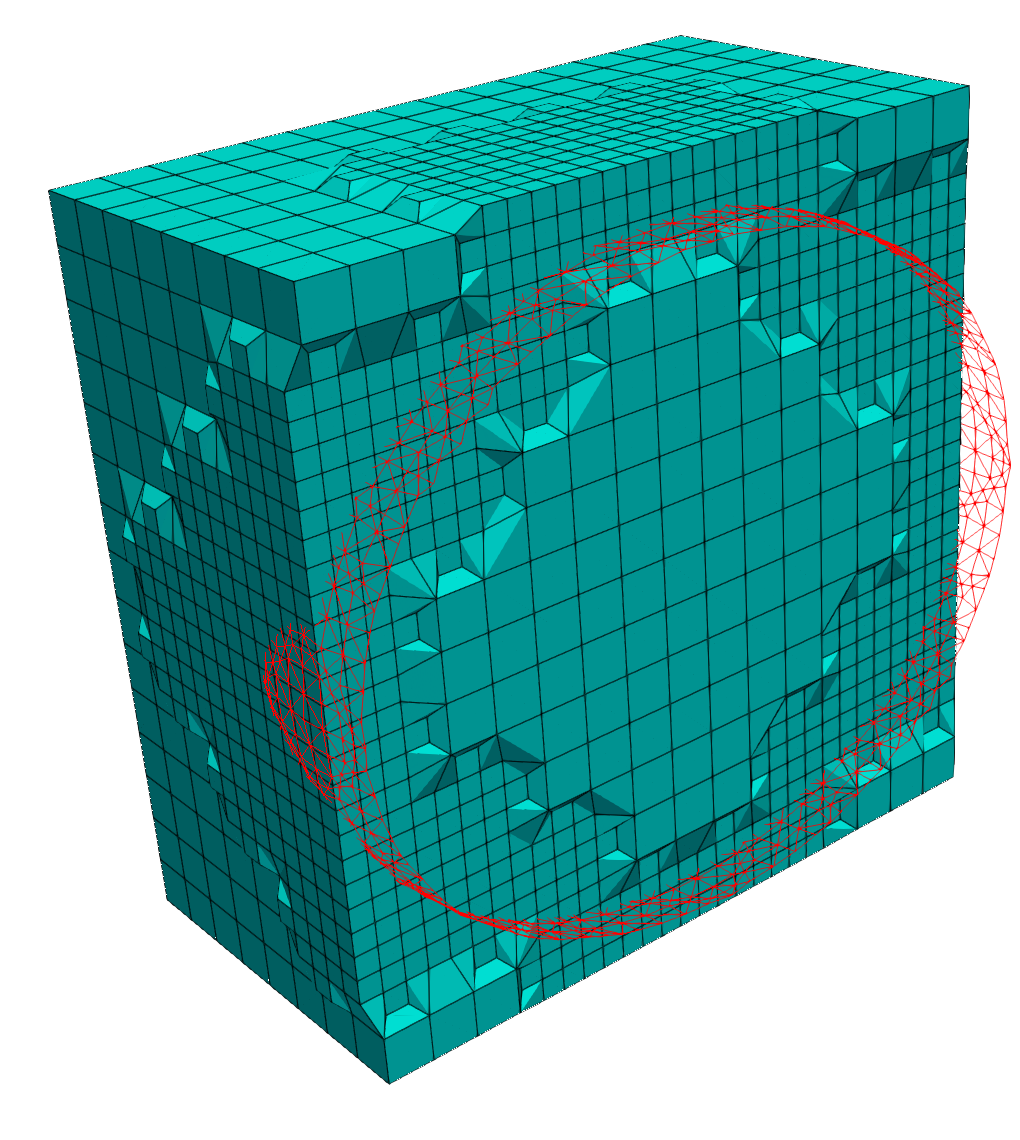}
    \label{fig:dualfull}  
    }
  \subfigure[]{
    \includegraphics[width=0.49\textwidth]{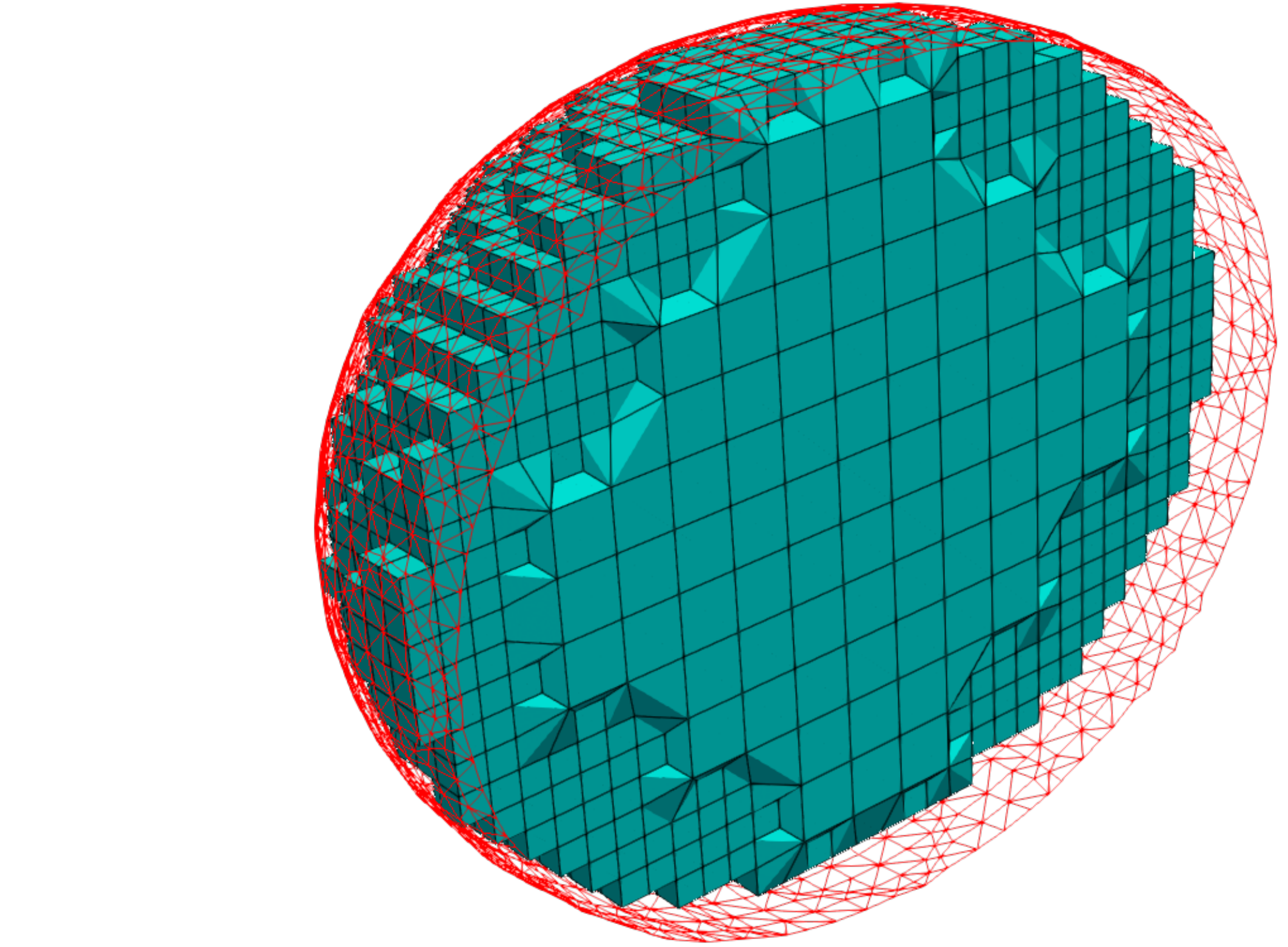}
    \label{fig:dual}  
  }\hspace{-4mm}
  \subfigure[]{
    \includegraphics[width=0.49\textwidth]{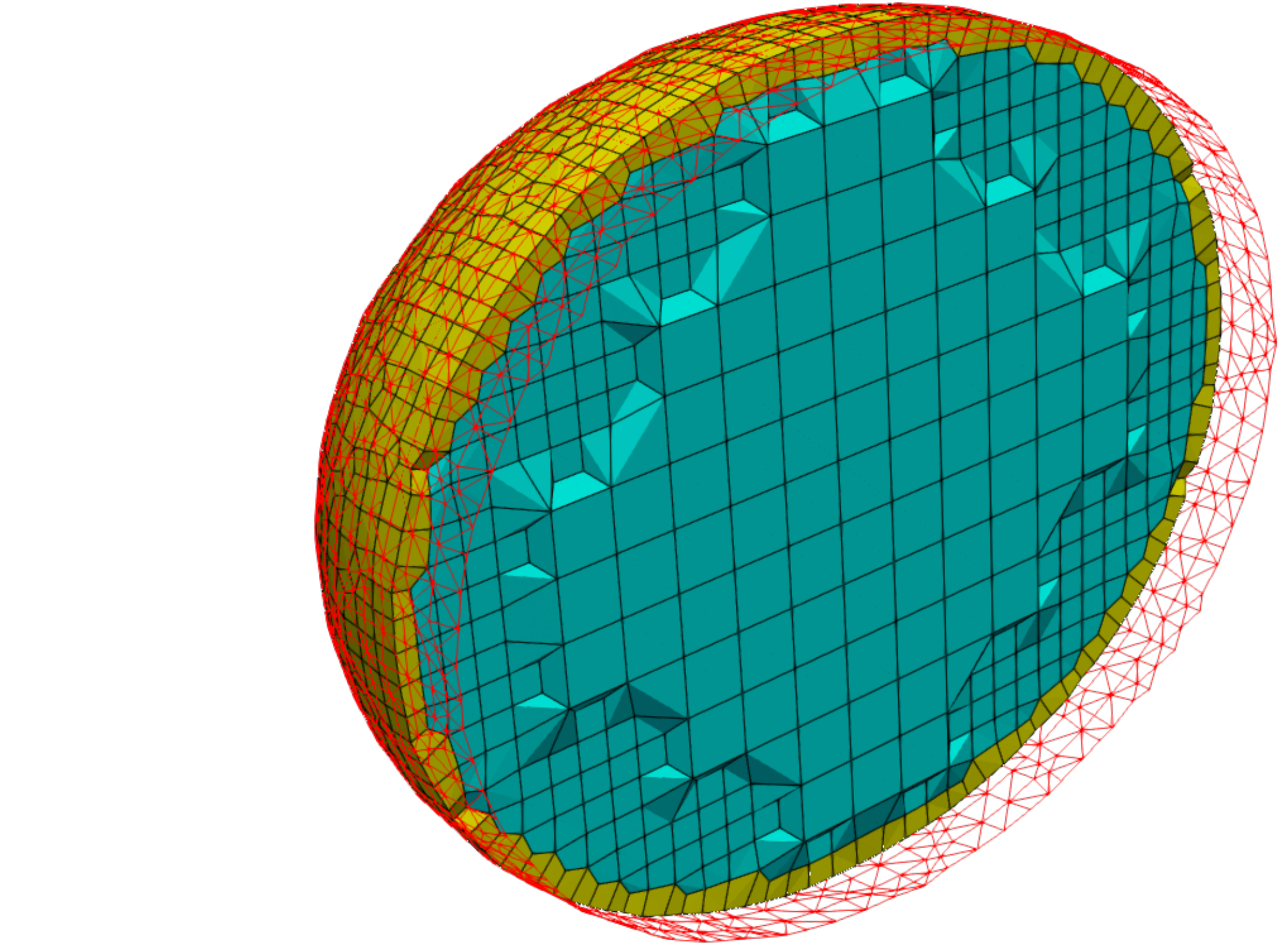}
    \label{fig:proj}  
    }
    \vspace{-4mm}
  \caption{HybridOctree\_Hex overview. (a) Initializing the octree from the surface triangular mesh (red) with feature preservation; (b) Transforming the initialized octree into a strongly balanced octree, satisfying the balancing rule and the pairing rule; (c) Constructing an all-hex dual mesh using pre-defined templates; (d) Clearing elements outside and around the boundary; (e) Meshing the buffer zone (yellow) with geometric fitting and Jacobian control.}
  \label{fig:bigpicture}  
\end{figure}

Starting from a closed, manifold surface triangular mesh that defines the domain to mesh, our goal is to generate high-quality adaptive all-hex meshes based on a hybrid octree. The detailed algorithm can be explained in five steps. As shown in Figure \ref{fig:bigpicture}, we first initialize an octree structure refined at high-curvature and narrow regions to capture detailed features. The octree is further refined to be strongly balanced following the balancing and pairing rules. Inspired by our earlier preliminary hybrid octree work \cite{hu2013adaptive}, we design templates for transition configurations to enable an all-hex dual mesh extraction from the strongly-balanced octree. After that, the elements outside the boundary and in the buffer zone are cleared. The buffer zone is then meshed by connecting the outmost points of the core mesh with their closest points on the input surface. In the final step, the smart Laplacian smoothing and optimization are coupled to improve the mesh quality with geometric fitting and minimum scaled Jacobian control.

\subsection{Initializing the Octree and Building A Strongly Balanced Octree Structure}
\label{subsec:2.1}

The input surface triangular mesh describes the boundary surface with a set of triangular elements. The triangular mesh is embedded into a cube, which is the root of the octree, marked as level $0$. Then, the octants are recursively subdivided based on the surface feature. Numerical simulations aim to use a minimal number of elements, while still preserving crucial features on the boundary surface, to ensure both efficiency and accuracy in calculations. To better capture high curvatures and narrow regions on the surface, we need to refine these regions to desired octree levels.

The Gaussian curvature $G$ at point $P_i$ is computed by $\lvert\lvert \sum_{j\in\mathcal{N}(i)}(\cot{\alpha_{ij}}+\cot{\beta_{ij}})(P_j-P_i)\rvert\rvert_2/(4A_i)$, where $j\in\mathcal{N}(i)$ is the index of vertices directly adjacent to $P_i$, $\alpha_{ij}$ and $\beta_{ij}$ are two angles opposite to edge $P_iP_j$, and $A_i$ is the Voronoi cell area around $P_i$ \cite{meyer2003discrete}. Five curvature thresholds $G_{thres} = \{0.5, 1, 2, 4, 8\}$ are adopted. If an octree cell at level $l\ (l = 0, 1, \cdots, 4)$ containing or intersecting with the triangular boundary  \cite{guigue2003fast} satisfies $G(l) > G_{thres}[l]$, we refine it to level $l + 5$. We use curvature to preserve detailed surface features. In addition, the thickness $T$ at point $P_i$ is measured by shooting a ray from $P_i$ along its normal direction to find the shortest intersection with other triangular elements on the boundary. We set five thickness thresholds $T_{thres} = \{16, 8, 4, 2, 1\}$, and refine all the octree cells with $T(l) < T_{thres}[l]$ to level $l + 5$. An octree structure is initialized after these operations, where the resulting octree level of each octant ranges from level $5$ to $9$. The following two rules are applied to convert the initialized octree to a strongly balanced one \cite{hu2013adaptive}.

\vspace{0.1in}
\noindent
\textbf{Balancing rule:} \textit{The level difference between two neighboring octants is at most one.}\vspace{0.1in}\\
\textbf{Pairing rule:} \textit{If an octant is subdivided to comply to the balancing rule, its siblings (the other seven octants belonging to the same parent) are subdivided along with it.}
\vspace{0.1in}

After implementing the balancing and pairing rules, the resulting strongly balanced octree consists of cubes with hanging nodes scattered within the octree structure due to the size variation. In this paper, we design templates based on the cutting procedure algorithm in our earlier preliminary work \cite{hu2013adaptive}, and the all-hex dual mesh is constructed using these templates. In the following, we provide key definitions pertinent to the subsequent algorithmic description.

\vspace{0.1in}
\noindent
\textbf{Hanging node:} \textit{A hanging node is a corner vertex hanging in the middle of neighboring faces or edges. In other words, it is a corner vertex for some elements but not for other adjacent elements containing it.}\vspace{0.1in}\\
\textbf{Octree (Quadtree) block:} \textit{In a 3D octree, eight children cells belonging to the same parent cell form an octree block. Similarly in a 2D quadtree, four children cells belonging to the same parent cell form a quadtree block.}\vspace{0.1in}\\
\textbf{Transition face:} \textit{A transition face is a face in an octree block that is shared by two octree cells with different octree levels.}\vspace{0.1in}\\
\textbf{Transition edge:} \textit{A transition edge is an edge in an octree block that is shared by three octree cells with different octree levels.}\vspace{0.1in}\\
\textbf{Hybrid octree:} \textit{A hybrid octree is derived from a strongly balanced octree to remove all hanging nodes and ensure each interior grid point is always shared by eight leaf cells \cite{hu2013adaptive}. The leaf cells of the hybrid octree can be polyhedra instead of all-hex.}

\subsection{Designing Templates and Constructing All-Hex Dual Mesh}
\label{subsec:2.2}

The strongly balanced octree exhibits gradual size variations, although it still contains hanging nodes throughout. A cutting algorithm, designed in \cite{hu2013adaptive}, aims to eliminate hanging nodes in both 2D quadtrees and 3D octrees. This algorithm strategically cuts strongly balanced quadtree/octree blocks into polygonal/polyhedral cells and extracts the dual mesh from the resulting hybrid quadtree/octree to obtain an all-quad/hex dual mesh. In 2D, the process involves cutting along the transition edge, while in 3D, it's more intricate, addressing five transition cases on faces and edges. The key point of the cutting method is to ensure that within the hybrid quadtree/octree derived from the strongly balanced quadtree/octree via the cutting procedure, every grid point is shared by four polygonal cells in 2D and eight polyhedral cells in 3D. This results in an all-quad/hex dual mesh generation. In this paper, we combine the two steps of creating a hybrid octree from a strongly balanced octree and generating a dual mesh of the hybrid octree together in our implementation by designing templates. We develop five templates in 3D to construct all-hex dual mesh directly by detecting transition faces and edges in the strongly balanced octree, which skips the sophisticated implementation of a hybrid octree and accelerates the entire meshing process. In 2D, only one template is needed. 

\vspace{0.1in}

\begin{figure}
  \centering
  \includegraphics[width=\textwidth]{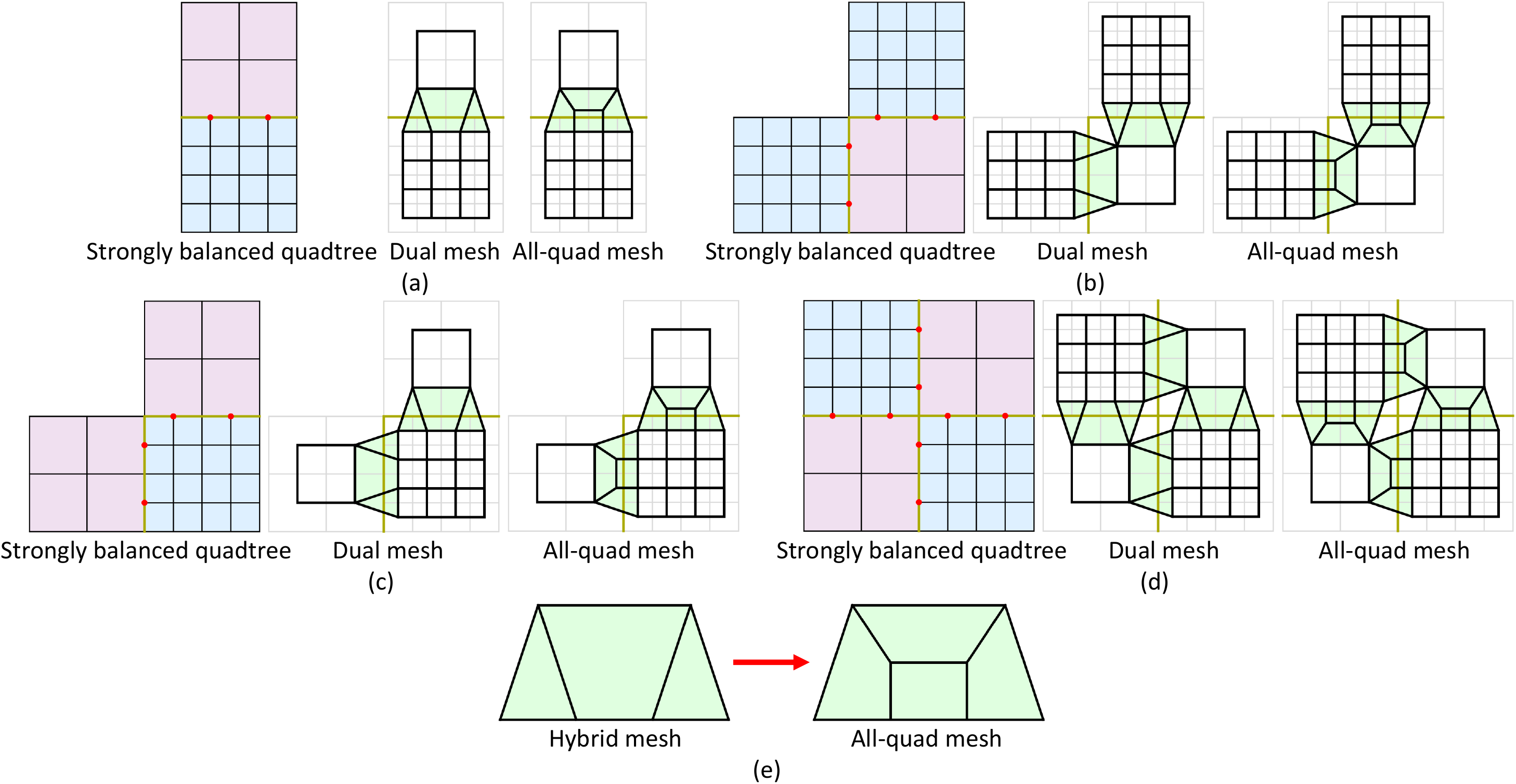}
  \vspace{-7.5mm}
  \caption{Quadtree transition configurations (a-d) showing the strongly balanced quadtree, hybrid dual mesh, and the resulting all-quad mesh after applying the transformation template in (e). There are two red hanging nodes on each golden transition edge. (a) Two neighboring quadtree blocks with one being refined; (b, c) Three neighboring quadtree blocks in an $\textit{L}$ shape with the left and right blocks refined or the corner block refined,  respectively; (d) Four neighboring quadtree blocks with a pair of diagonal blocks refined.}
  \label{fig:transition2D}  
\end{figure}

\textbf{2D template.} In a strong balanced quadtree, four quadrants belonging to the same parent form a quadtree block. A transition edge is formed when two neighboring quadtree blocks have different levels. Four possible configurations occur in 2D. In Figure \ref{fig:transition2D}(a), when one of the two adjacent quadtree blocks (blue) is subdivided, there is a transition edge (golden) between the pink and blue blocks. There are two hanging nodes on the edge, marked by red dots. Note that each hanging node is shared by three cells while each regular grid point is shared by four cells. To obtain the dual mesh of the strongly balanced quadtree \cite{zhang2006adaptive, zhang2009surface}, each cell center is selected as the associated mesh vertex. For each grid point, its dual element is created by connecting the mesh vertices from cells sharing it. Given that every grid point is shared by either three or four cells, the resulting dual mesh consists of triangles and quads.

The resulting triangles in the dual mesh need to be removed. We design a template in Figure \ref{fig:transition2D}(e) to transform those two triangles and one quad in between into four quads without affecting the surrounding connectivity. Furthermore, when transitions occur in different directions to the same quadtree block, we transform them independently with the help of the pairing rule. Figure \ref{fig:transition2D}(b-d) show configurations with both horizontal and vertical transition edges, and each direction is similar to Figure \ref{fig:transition2D}(a). 
By checking all the possible configurations, we obtain one unique transition template as shown in Figure \ref{fig:transition2D}(e) to achieve all-quad dual mesh extraction from the strongly-balanced quadtree.

\begin{lemma}
\label{lem:quad}
For dual mesh extraction from the strongly balanced quadtree, only one transition template is needed: the hybrid elements consisting of two triangles and one quad in between are converted to four quads, as shown in Figure \ref{fig:transition2D}(e).
\end{lemma}

\noindent
 \textbf{Discussion 2.1.} In the case of a strongly-balanced quadtree, we directly detect the transition edges and generate all-quad dual meshes using the designed template. This eliminates the need to generate a complex and memory-intensive hybrid quadtree, leading to significant improvements in computational efficiency. By independently addressing each transition edge with the dedicated template, we ensure a seamless mesh transformation process.
 \vspace{0.1in}

\begin{figure}
  \centering
  \includegraphics[width=\textwidth]{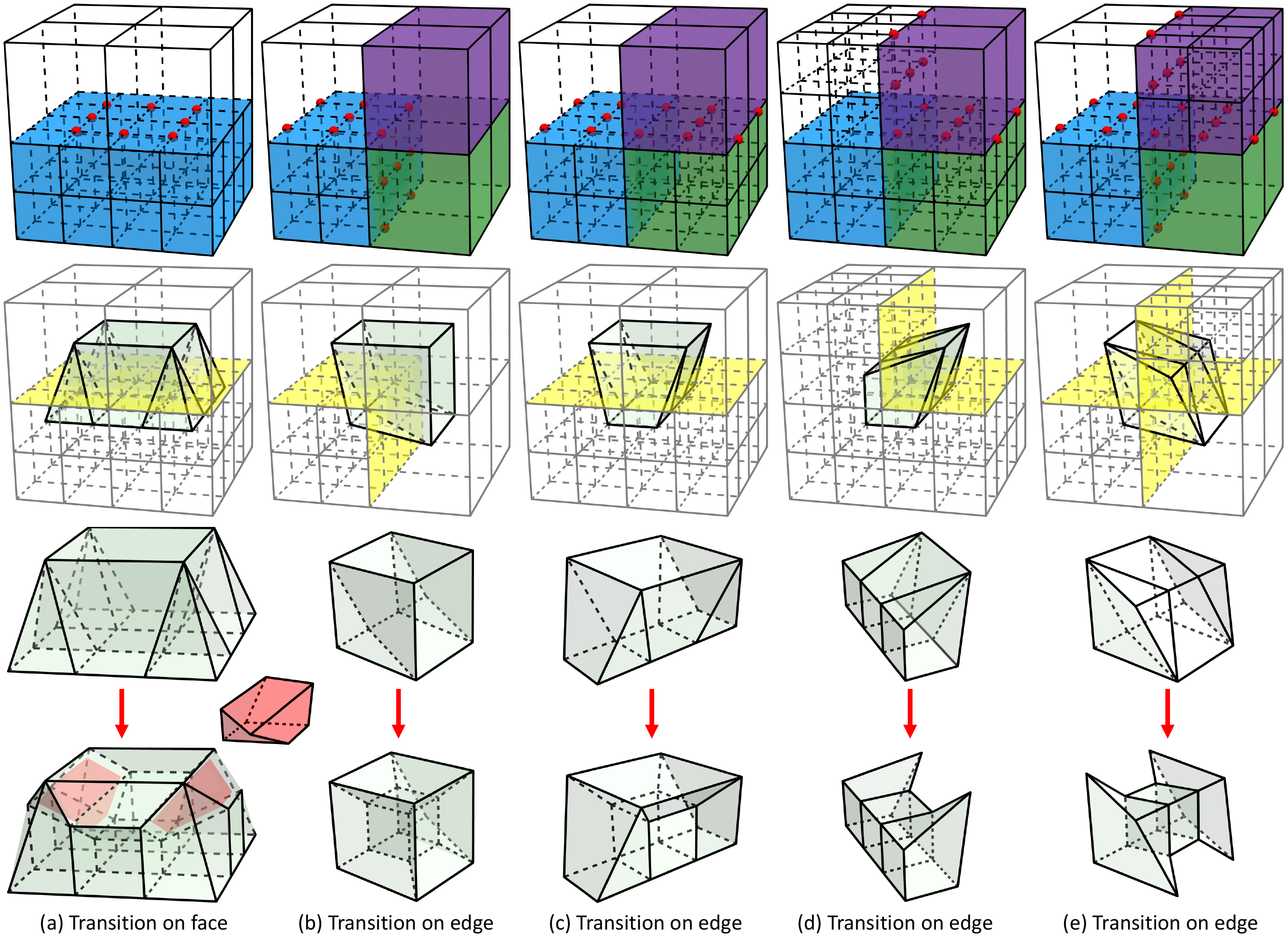}
  \vspace{-7.5mm}
  \caption{Octree transition configurations (a-e) showing the strongly balanced octree (first row), hybrid dual mesh (second row), and the transformation template (third and fourth row). Hanging nodes are in red, transition faces are in yellow, and octants with the same color belong to the same block. (a) Two neighboring octree blocks with one being refined, resulting in the transition on the face. The hexes with the minimum scaled Jacobian $0.258$ among all the templates are highlighted and extracted; (b-e) Four neighboring octree blocks at different levels sharing a transition edge.}
  \label{fig:frustum}  
\end{figure}

\textbf{3D templates.} The 3D transition configurations are similar to the 2D cases, albeit with significantly greater complexity \cite{hu2013adaptive}. Following the 2D cases, we check all the possible configurations and design a template to transform each configuration independently. Based on the pairing rule, an octree block is created by grouping eight child octants that originate from the same parent octant. This grouping occurs because these child octants have identical behavior - either all of them are further subdivided, or none of them is subdivided. A transitional scenario arises when the adjacent blocks have level differences. Degenerated elements appear on the transition faces/edges, necessitating their conversion into all-hex elements. Similar to 2D, the face transition case is shown in Figure \ref{fig:frustum}(a). Octants of the same color (blue or pink) belong to the same block. Only half of each block adjacent to the transition face is depicted. There is a yellow transition face between these two blocks since the blue block is subdivided, the red dots represent hanging nodes on the transition face, and the dual mesh of the strongly balanced octree is shaded in light green, consisting of one hex, four pyramids, and four triangular prisms. The designed template can generate an all-hex dual mesh to deal with the face transition scenario. Unlike 2D transitions, 3D transitions maintain quad elements on the top and bottom faces of the frustum but introduce two new vertices on each side face. Therefore, we must enumerate all transition cases involving neighboring octree blocks and devise a transitional template for each to accommodate the two new vertices generated by the face transition template.

By checking all the possible configurations, the position relationship between two transition faces falls into four distinct cases, corresponding to four octree blocks sharing a transition edge at different levels. One, two, and three of the four octree blocks are subdivided to a finer level as shown in Figure \ref{fig:frustum}(b-e). Only one-fourth of each block adjacent to the transition edge is drawn, octants with the same color (blue, green, or purple) belong to the same block. Figure \ref{fig:frustum}(b, d) are the configurations between two vertically placed face transitions. Two new points are placed at the diagonal to match face transition templates. Figure \ref{fig:frustum}(c) depicts the configuration between two face transitions placed side by side. Two new points are placed on each side face to match the face transition. Figure \ref{fig:frustum}(e) is a special case of Figure \ref{fig:frustum}(b, d), where four face transitions are placed in a cross. Two new points are placed at the two diagonals to match the face transitions on both diagonals. For each transition case, the number of polyhedra and the resulting hexes after transformation is listed in Table \ref{tab:3D}. The number of polyhedra and hexes in Figure \ref{fig:frustum}(a) is greater than the other configurations because Figure \ref{fig:frustum}(a) is the face transition case, and the other configurations are edge transition cases.

\begin{table}
\caption{Number of polyhedra and hexes in each template}
\vspace{-3.5mm}
\label{tab:3D}
\small
\centering
\begin{tabular}{cccccc}
\toprule
\multicolumn{1}{c|}{\multirow{2}{*}{Element Type}} & \multicolumn{1}{c|}{Transition on Face} & \multicolumn{4}{c}{Transition on Edge} \\
\cline{2-6}
\multicolumn{1}{c|}{} & \multicolumn{1}{c|}{Figure \ref{fig:frustum}(a)} & Figure \ref{fig:frustum}(b) & Figure \ref{fig:frustum}(c) & Figure \ref{fig:frustum}(d) & Figure \ref{fig:frustum}(e) \\
\hline
\multicolumn{1}{c|}{Polyhedra} & \multicolumn{1}{c|}{9} & 3 & 3 & 3 & 3 \\
\multicolumn{1}{c|}{Hexes} & \multicolumn{1}{c|}{13} & 5 & 4 & 3 & 3 \\
\bottomrule
\end{tabular}
\end{table}

It is worth noting that when an octree block has three transition faces in the XYZ directions, our templates leave a hex at the corners of the transition, which is not shown in Figure \ref{fig:frustum}. By placing hexes at those corners, we achieve an all-hex dual mesh. Additionally, we have a minimum scaled Jacobian of $0.258$ in the designed templates which will be further improved in the resulting dual mesh through quality improvement.

\begin{lemma}
\label{lem:hex}
For dual mesh extraction from the strongly balanced octree, five transformation templates are specifically designed to handle one face transition and, additionally, four edge transitions. These templates are based on the resulting dual mesh of the hybrid octree described in \cite{hu2013adaptive}. The cutting procedure ensures that each grid point in the obtained hybrid octree is always shared by eight polyhedra, resulting in an all-hex dual mesh. Therefore after applying the transition templates, our resulting mesh is guaranteed to be all-hex.
\end{lemma}

\noindent
\textbf{Discussion 2.2.} In contrast to the cutting method described in \cite{hu2013adaptive}, our approach not only produces the same all-hex meshes but also introduces template-based transformations specifically tailored for independently handling transition edges and faces. The cutting method involves complex polygonal/polyhedral cells in the hybrid quadtree/octree, which are challenging to store and manipulate during software implementation. Instead, our octree transformation employs five pre-defined templates to independently address each transition face or edge, simplifying the mesh generation process and enhancing the ease of implementation. Specifically, we directly detect transition faces and edges in the strongly balanced octree and apply these pre-defined templates to generate an all-hex mesh. This approach obviates the need for a complex and memory-intensive hybrid octree, thereby significantly improving computational efficiency.

\subsection{Clearing Buffer Zone}
\label{subsec:2.3}

After generating the all-hex dual mesh, the interior core mesh is obtained by removing elements outside and in the vicinity of the boundary surface. The region near the boundary is referred to as the buffer zone. Similar to \cite{zhang2013robust}, if the minimum distance from a vertex to the boundary falls below a threshold $\epsilon_s = s_{max}/2$, where $s_{max}$ represents the maximum size among the elements sharing that vertex, all elements connected to it are deleted. During the implementation, we observed that this setting can be sensitive to large elements located in size transition regions, potentially leaving holes in the surface. To address this issue, we calculate the signed distance function for corner points associated with every hex element \cite{paragios2002matching, rousson2002shape}. Each hex has eight signed distance functions $f(x^i)$, where $i = 0,\ 1,\ \cdots,\ 7$, attached to the eight corner points. We compute $f_{min}$ and $f_{max}$ and remove the hex element if the condition $f_{min} + 0.1 \times f_{max} < 0$ is met.

A buffer zone exists between the interior core mesh and the input surface. 
The extracted core mesh cannot be used directly for buffer layer meshing, and we need a more precise operation to remove elements on the boundary of the core mesh that may lead to poor-quality elements in the buffer zone. We adopt the scaled Jacobian as our metric for this purpose \cite{zhang2006adaptive}. Within each hex, for every corner node $x$, three edge vectors are defined as $e_i=x_i-x$ $(i = 0,\ 1,\ 2)$. Then the Jacobian matrix at $x$ is defined as $[e_0, e_1, e_2]$, and its \textit{Jacobian} $J(x)$ is the determinant of the Jacobian matrix. We obtain the \textit{Scaled Jacobian} $\mathit{SJ}(x)$ if $e_0,\ e_1,\ e_2$ are normalized. For each hex, we compute the (scaled) Jacobian at eight corners and the body center. For the body center, $e_i$ ($i$ = 0, 1, 2) is computed using three pairs of opposite face centers.

Here we introduce a restriction on the buffer zone clearance. Assume $x$ is a boundary point of the core mesh, and it is shared by $m$ quad faces on the core mesh boundary. We denote the normal vectors of triangles formed by $x$ and two adjacent points in all the quad faces as $n_i$, where $i = 0,\ 1,\ \cdots,\ m - 1$. This represents that $m$ new hexes will be generated around $x$ after meshing the buffer zone. These hexes are denoted as $h_i, i = 0,\ 1,\ \cdots,\ m - 1$. The edge vector from $x$ to its corresponding surface point is denoted as $e$. According to the scaled Jacobian definition, a valid $e$ needs to satisfy $\min{\mathit{SJ}(h_i)} \leq \min{\mathit{SJ}(x)} \leq \min (n_i\cdot e)$, where $i = 0,\ 1,\ \cdots,\ m - 1$. If $\min(n_i\cdot e) < 0$, then we have $\min{\mathit{SJ}(h_i)} < 0$. Therefore we define the restriction: 

\vspace{2mm}
\noindent \textbf{Restriction for buffer zone clearance:} \textit{Any three normals $n_i$, $n_j$ and $n_k$ of the $m$ quad faces surrounding $x$ need to satisfy $(n_i \times n_j) \cdot n_k > 0$. In the implementation, we remove one hex attached to $x$ iteratively until $(n_i \times n_j) \cdot n_k > 0$ for all $i,\ j,\ k = 0,\ 1,\ \cdots,\ m - 1$ and $i \neq j \neq k$.} 
\vspace{2mm}

\begin{figure}
  \centering
    \subfigure[]{
    \includegraphics[width=0.4\textwidth]{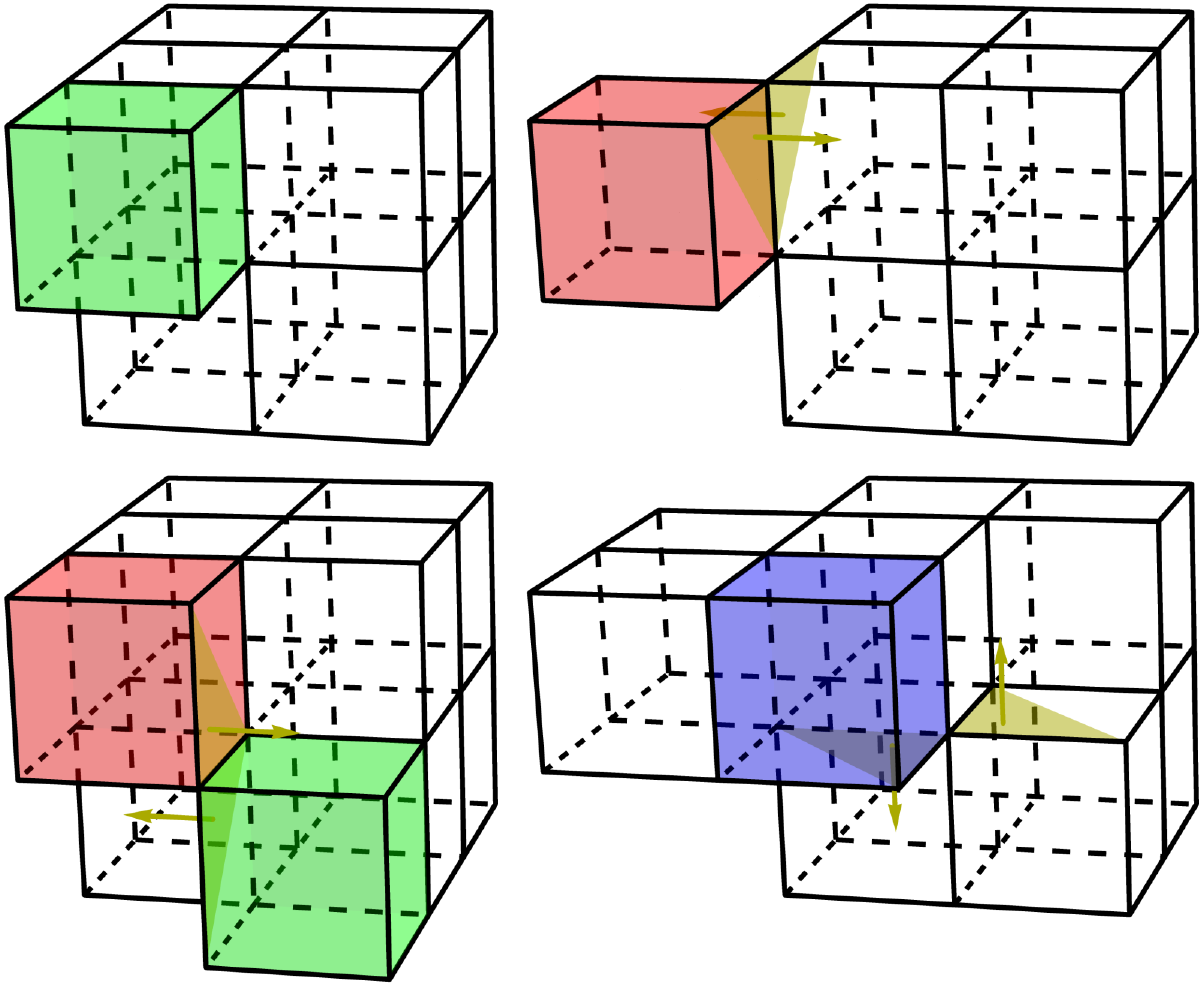}
    \label{fig:defect1}
    }\hspace{10mm}
  \subfigure[]{
    \includegraphics[width=0.32\textwidth]{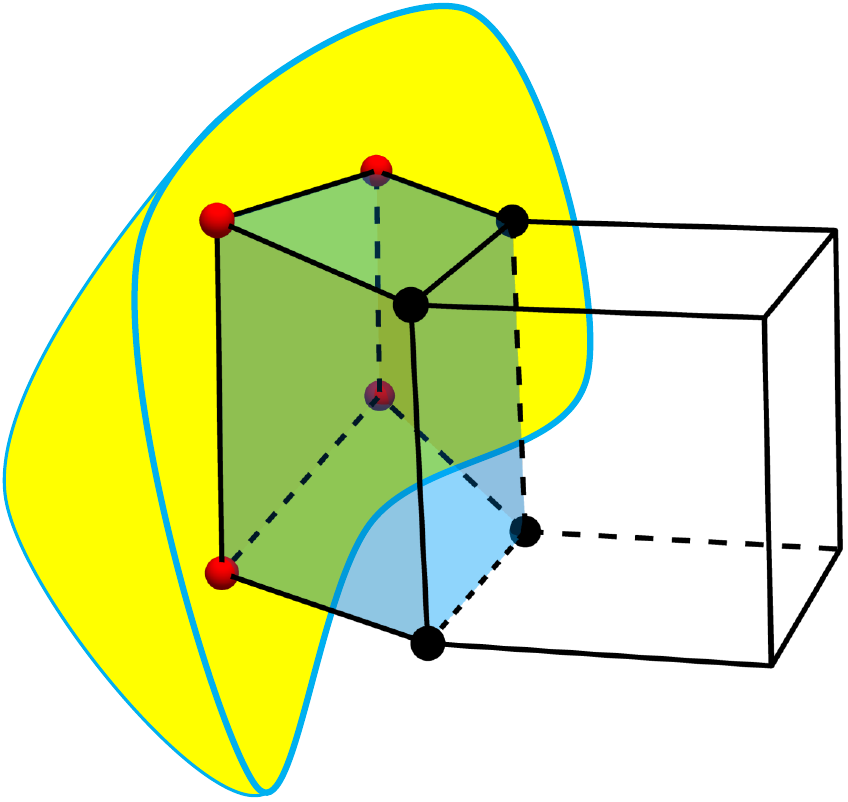}
    \label{fig:defect2}  
    }
  \vspace{-4.5mm}
  \caption{(a) Four cases comparing with \cite{zhang2013robust}.
  Hexes removed by both methods are in red, hexes removed only by \cite{zhang2013robust} are in green, and the hex removed by our method only is in blue. Shaded yellow triangles represent their normals violating our restriction and one adjacent hex needs to be removed; (b) The buffer layer is formed by connecting core mesh boundary points (black dots) and their corresponding points (red dots) on the input surface.}
  \label{fig:defect}  
\end{figure}

As shown in Figure \ref{fig:defect}(a), our buffer zone clearance restriction is different from \cite{zhang2013robust}, which deletes the single elements sharing only a point, an edge, or a face with other elements, as well as elements with non-manifold connectivity. We notice that the non-manifold criterion is not sufficient to ensure good element quality in the buffer zone. Instead, the normals around a boundary point of the core mesh are the key factors, and we believe this is the main reason why we could improve the buffer zone element quality to surpass the interior template element quality (the minimum scaled Jacobian of the template elements is $0.258$, which will be improved to higher than $0.5$ after quality improvement; see Section \ref{subsec:2.4}).

Finally, it is worth mentioning that we iterate over all the surface points of the core mesh. For each surface point, we iteratively remove one hex in each iteration until all the surface points satisfy the restriction. In each iteration, we record the hexes that are attached to boundary points and that violate the restriction, along with the number of boundary faces they share. We prioritize deleting the hex with the most boundary faces to prevent leaving holes in the resulting core mesh. 

\subsection{Quality Improvement with Jacobian Control}
\label{subsec:2.4}

In the last step, we mesh the buffer zone by connecting each boundary point $x_i$ of the core mesh with its closest point on the input triangular surface $x_i^s$ to form a hex element $h_i$, as shown in Figure \ref{fig:defect}(b). The resulting elements in the buffer layer may exhibit poor quality or even possess a negative Jacobian. The objective is to preserve the positions of boundary points on the surface while relocating corners of worst-Jacobian elements to enhance the overall mesh quality. Here, we couple optimization with smart Laplacian smoothing. The optimization is performed in each iteration to identify the worst-Jacobian element in the entire mesh and adjust its corner points, whereas smart Laplacian smoothing is performed every $1,000$ iterations on the outmost two layers of vertices only to expedite the quality improvement. Specifically, within the smart Laplacian smoothing procedure, for surface points, we compute the average position of their neighboring surface points and determine the closest point on the input surface to this average position; for interior points, we calculate the average of their neighboring points. We only relocate a point if the minimum scaled Jacobian of the elements affected by the movement exceeds $\epsilon_{\mathit{SJ}}$ or if the distance from the point to the input surface is the greatest.

In the optimization, we propose a new energy function, consisting of the geometry fitting, scaled Jacobian and Jacobian terms. We have
\begin{equation}
\label{equ:1}
E = E_{\mathit{GF}} - E_{\mathit{SJ}} - E_J = \sum_{i=0}^{\mathit{nvert} - 1} \lVert x_i - x_i^s \rVert_2^2 - \sum_{i = 0}^{m - 1} \min\mathit{SJ}(h_i) - \sum_{i = 0}^{n - 1} \min J(h_i),
\end{equation}
where $\mathit{nvert}$ is the number of surface vertices, $m$ is the number of positive-Jacobian hexes, and $n$ is the number of negative-Jacobian hexes. We adopt a gradient-based method to iteratively minimize the energy function. All the mesh vertices are optimized by
\begin{equation}  
x_i \rightarrow x_i - \alpha\nabla E |_{x_i}, \quad i = 0, 1, \dots, 2\times\mathit{nvert} - 1,
\end{equation}
where we choose the weight $\alpha = 0.8 \times 10^{-3}$ for all the tested models in this paper.

In finding the closest surface point $x_i^s$ for each boundary point $x_i$ of the core mesh efficiently,  we iterate through all the triangles in a bounding box with the size of 10 times the local triangular edge length and compute the closest point on each triangle to $x_i$. 
For faster computation, the closest triangle number to each $x_i$ is updated every $1,000$ iterations. If the maximum value of the minimum distance from all points to the surface is $< 10^{-6}$, we directly pull every $x_i$ onto its corresponding $x^s$.

\begin{figure}
  \centering
  \subfigure[]{
    \includegraphics[width=0.21\textwidth]{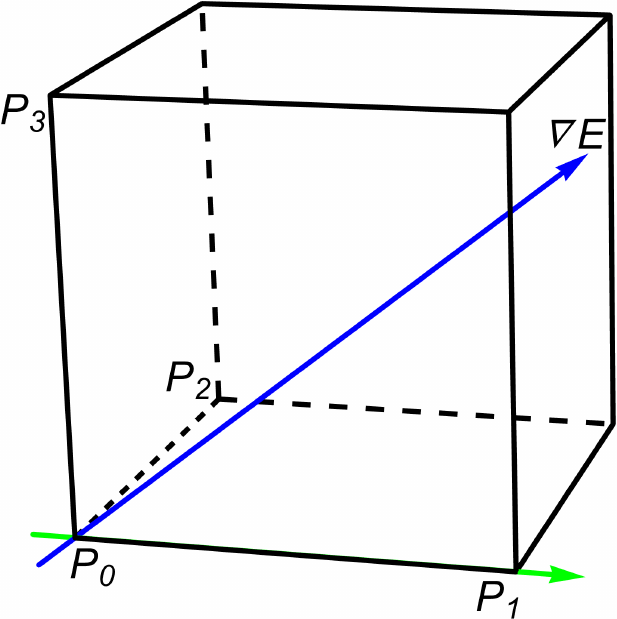}
    \label{fig:simplecube}  
  }
  \subfigure[]{
    \includegraphics[width=0.37\textwidth]{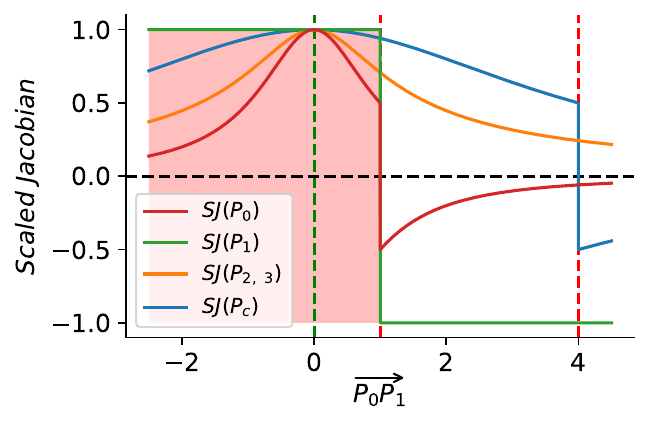}
    \label{fig:sjaj}  
    }\hspace{-2mm}
  \subfigure[]{
    \includegraphics[width=0.37\textwidth]{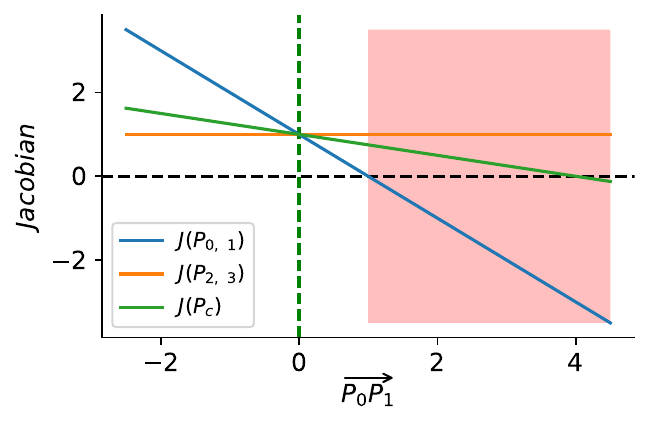}
    \label{fig:sjaj2}  
    }\vspace{-3mm}
  \subfigure[]{
    \includegraphics[width=0.37\textwidth]{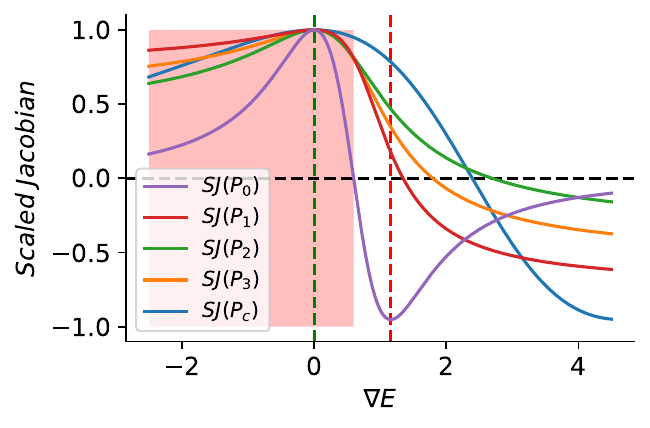}
    \label{fig:sjaj3}  
    }
  \subfigure[]{
    \includegraphics[width=0.37\textwidth]{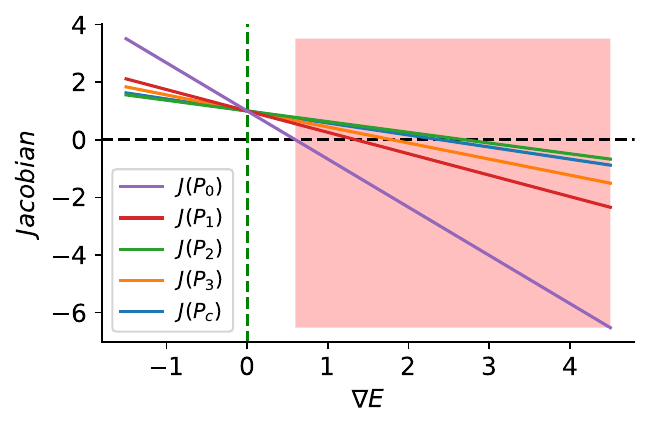}
    \label{fig:sjaj4}  
    }
  \vspace{-4.5mm}
  \caption{(a) A hex with $|P_0P_1|=1$ is to be optimized, where $P_0$ moves along the $\overrightarrow{P_0P_1}$ direction (green arrow) in (b, c) or an arbitrary direction $\nabla E$ (blue arrow) in (d, e). The optimum point is $|P_0P_1|$ or $|\nabla E| = 0$, marked as dashed green lines in (b-e). The intervals of adopting Jacobian or scaled Jacobian in the algorithm are shaded in red. The in-differentiable points and valleys are marked as dashed red lines in (b, d); (b, d) The scaled Jacobian at $P_0, P_1, P_2, P_3$, and the body center $P_c$; (c, e) The Jacobian curves.}
  \label{fig:sjproblem}  
\end{figure}

The scaled Jacobian and Jacobian terms in Equation (\ref{equ:1}) are only applied to hexes with the scaled Jacobian below a pre-defined threshold $\epsilon_{\mathit{SJ}}$, 
which is initialized as $0.01$. In every $1,000$ iterations, we increase $\epsilon_{\mathit{SJ}}$ by $0.01$ until all the hexes have $\min\mathit{SJ}(h_i) \geq \epsilon_{\mathit{SJ}}$ and all the boundary points are on the input surface, and finish the optimization if the mesh quality cannot be further improved. Note that when only using the scaled Jacobian in the optimization, one problem arises: the scaled Jacobian is in-differentiable at $|P_0P_1| = 1,\ 4$ as shown in Figure \ref{fig:sjproblem}(b) or has a valley at $|\nabla E| = 1.161$ as shown in Figure \ref{fig:sjproblem}(d). To address this issue, we use the combined scaled Jacobian and Jacobian approach. The scaled Jacobian term is selected for optimization when it is differentiable and positive. Otherwise, the Jacobian term is chosen, which is differentiable in the entire domain and does not have minima in the negative-Jacobian region. It worth noting that the Jacobian value cannot be used in the region of $\min J(h_i) > 0$, because the to-be-optimized point cannot converge to the optimal position $|P_0P_1|$ or $|\nabla E| = 0$ and will continue to $|P_0P_1|$ or $|\nabla E| \rightarrow -\infty$ as shown in Figure \ref{fig:sjproblem}(c, e).

 \vspace{0.1in}
\noindent
 \textbf{Discussion 2.3.} For quality improvement, previous methods first remove non-manifold points on the core mesh surface and then apply geometric flow to enhance the overall mesh quality \cite{zhang2006quality, zhang2009surface}. Afterward, optimization-based smoothing is employed to improve the lowest-quality elements of the mesh \cite{freitag1997combining}. In contrast to this, our approach involves computing face normal directions around each surface point. Subsequently, we remove one or multiple hexes, ensuring that no strict angle limitations exist during the meshing of the buffer zone. Within the gradient-based optimization, we utilize a combined scaled Jacobian and Jacobian method to avert the optimizer from getting stuck in local minima. Numerical results presented in Section \ref{sec:3} further indicate that our novel buffer zone clearance and mesh quality enhancement techniques lead to a significantly higher minimum scaled Jacobian ($>$ 0.5).

\section{Numerical Results and Discussion}
\label{sec:3}

\begin{figure}
  \centering
  \includegraphics[width=\textwidth]{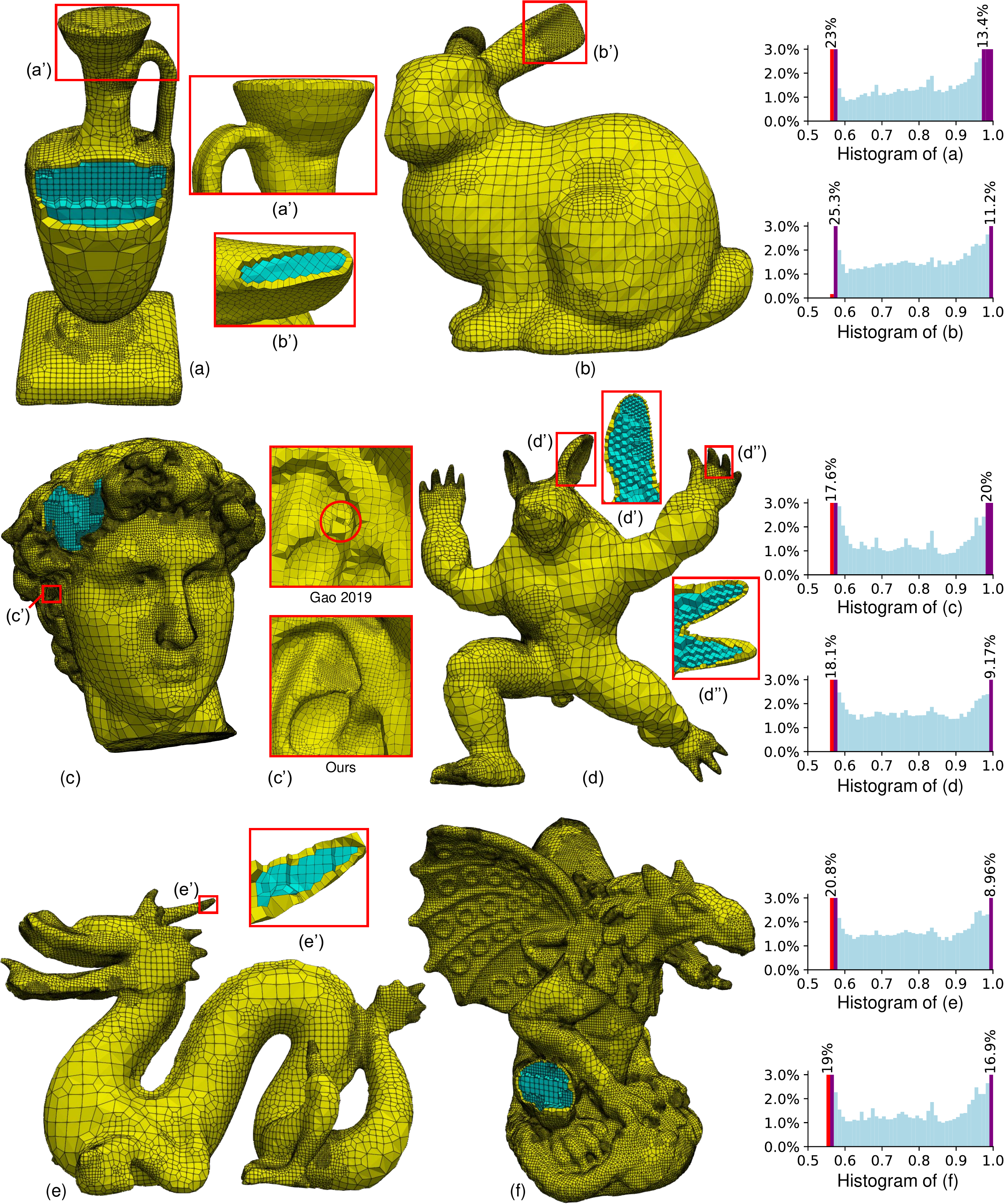}
  \vspace{-6.5mm}
  \caption{(a) The bottle1; (b) The bunny; (c) The david; (d) The deformed armadillo; (e) The dragon stand; and (f) The gargoyle. (a', b', c', d', d'', e') show zoomed-in images; (c') shows a zoomed-in comparison between our mesh and \cite{gao2019feature} mesh, which generates a hole at the thin region; and the scaled Jacobian histograms are shown in the last column. The red bar represents the minimum scaled Jacobian and purple bars are truncated ones due to a higher frequency ($\geq 3\%$).}
  \label{fig:mesh012345}  
\end{figure}

\begin{figure}
  \centering
  \includegraphics[width=\textwidth]{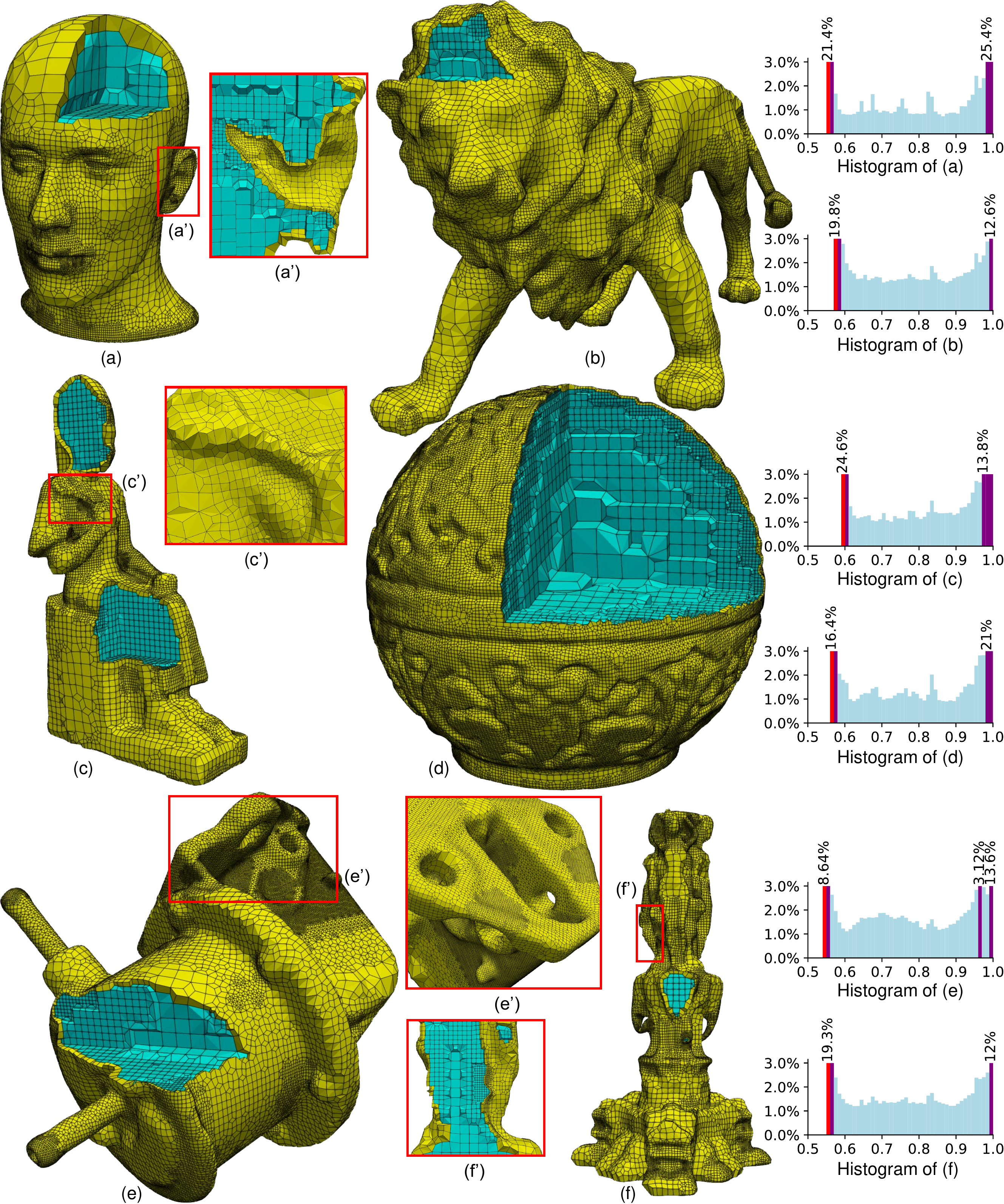}
  \vspace{-5mm}
  \caption{(a) The head; (b) The lion recon; (c) The ramses; (d) The red circular box; (e) The oil pump; and (f) The thai statue. (a', c', e', f') demonstrate zoomed-in images; and the scaled Jacobian histograms are shown in the last column. The red bar represents the minimum scaled Jacobian, and the purple bars are truncated ones due to a higher frequency ($\geq 3\%$).}
  \label{fig:mesh67891011}  
\end{figure}

We run HybridOctree\_Hex on a range of complex models without the need for parameter adjustments and compare our algorithm with previous methods \cite{hu2013adaptive, gao2019feature}. Twelve representative meshing outcomes are displayed in Figures \ref{fig:mesh012345} and \ref{fig:mesh67891011}. Our results were computed on a PC equipped with a 3.6 GHz Intel i7-12700 CPU and 32GB of memory.

The presented models demonstrate a range of intricate features such as high curvature and narrow regions, which are widely present in Figure \ref{fig:mesh012345}(b, d, e, f) through tips, as well as slim cylindrical structures in Figure \ref{fig:mesh67891011}(b, e, f). High-genus topologies can also lead to narrow regions; see Figure \ref{fig:mesh012345}(a, e) and Figure \ref{fig:mesh67891011}(e, f). If these narrow regions are not adequately detected and refined, the resulting mesh may exhibit holes, as exemplified in the zoom-in picture \cite{gao2019feature} of Figure \ref{fig:mesh012345}(c'). By incorporating the thickness measurement, our algorithm maintains the input topology faithfully. Note that the model's thickness bears a direct proportion to $cos \theta$, where $\theta$ is the angle between the thickness direction and the octree structure orientation. In the worst case following the long diagonal direction of the octree structure, we have $\theta = \arcsin(\sqrt{3}/3) \approx 35.3^\circ$ and more elements are needed to preserve the narrow region. In the oil pump model (Genus-4) shown in Figure \ref{fig:mesh67891011}(e), the octree orientation exhibits $\theta = 30^\circ$ with the thin plate's thickness direction. Our algorithm demonstrates its ability to preserve the model's topology correctly without leaving holes on the thin plates, although it requires a higher mesh density. Figures \ref{fig:mesh012345}(c-f) and \ref{fig:mesh67891011}(a-d, f) showcase a diverse array of detailed features. Without proper detection and refinement, these features may appear blurred in the final mesh. Our findings reveal that the curvature function is remarkably proficient in pinpointing such crucial surface details, and the quality improvement algorithm faithfully reproduces them in the resulting meshes. This is evident in the representation of eyes in Figures \ref{fig:mesh012345}(c, f) and \ref{fig:mesh67891011}(a), as well as noses in Figures \ref{fig:mesh012345}(c, d, f) and \ref{fig:mesh67891011}(a, b).

Table \ref{tab:mesh-statistics} presents a thorough analysis of mesh statistics. When considering mesh size alongside refinement level, our method demonstrates superior mesh adaptation. This is achieved by employing local refinement exclusively in regions of high curvature and narrow regions, resulting in meshes with elevated refinement levels that minimize redundant elements. Through the utilization of face-normal-based core mesh surface enhancement, coupled with the Jacobian and scaled Jacobian optimization technique, we have attained notably improved minimum scaled Jacobians ($> 0.5$). 
Additionally, our method excels in terms of efficiency, generating meshes in the shortest time in most models compared to \cite{hu2013adaptive, gao2019feature}. This is attributed to our strategic implementation of smart Laplacian smoothing, which is restricted to the outermost two layers of vertices and executed once every $1,000$ iterations. Previous methods often employ this technique for every mesh vertex on a global scale.

\begin{table}
\caption{Mesh statistics of the testing models}    
\label{tab:mesh-statistics} 
\small
\vspace{-3.5mm}
\centering 
\begin{tabular}{cccccc}    
\toprule
\vspace{-1.5mm} \multirow{1.6}{*}{{Model}} & \multirow{1.6}{*}{Method} & Mesh Size & Scaled Jacobian & Refinement & Time \\    
& & \scriptsize{(vertex $\sharp$; element $\sharp$)} & \scriptsize{[worst; best]} & \multirow{1.3}{*}{Level}  & \scriptsize{(s)}\\
\midrule
Bottle1 & \cite{gao2019feature} & (39,326; 33,635) & [0.181; 0.999] & 4 & 8,175\\
(Genus-1) & ours & \textbf{(36,091; 30,145)} & \textbf{[0.560; 1.0]} & 4 & \textbf{218}\\
\midrule
\multirow{3}{*}{{Bunny}} & \cite{hu2013adaptive} & (46,476; 39,605) & [0.00100; 1.0] & 3 & 462\\
& \cite{gao2019feature} & (35,330; 29,698) & [0.292; 0.999] & 4 & 3,569\\
(Genus-0)& \cite{zhang2013robust} & (119,799; 106,730) & [$3.85\times10^{-5}$; 1.0] & 3 & 258\\
& ours & \textbf{(26,375; 21,695)} & \textbf{[0.570; 1.0]} & 4 & \textbf{358}\\
\midrule
David & \cite{gao2019feature} & \textbf{(127,778; 112,314)} & [0.126; 0.999] & 4 & 38,866\\
(Genus-0)& ours & (319,465; 282,957) & \textbf{[0.560; 1.0]}& 6 & \textbf{10,450}\\
\midrule
Deformed Armadillo & \cite{gao2019feature} & (43,591; 35,611) & [0.0678; 0.999] & 4 & 6,573\\
(Genus-0) & ours & \textbf{(43,216; 34,939)} & \textbf{[0.560; 1.0]} & 5 & \textbf{3,431}\\
\midrule
Dragon Stand2 & \cite{gao2019feature} & (74,618; 61,441) & [0.0290; 0.999] & 5 & 23,062\\
(Genus-1)& ours & \textbf{(62,576; 50,853)} & \textbf{[0.560; 1.0]} & 4 & \textbf{2,052}\\
\midrule
Gargoyle & \cite{gao2019feature} & \textbf{(157,008; 135,737)} & [0.0702; 0.999] & 4 & ---\\
(Genus-0)& ours & (273,704; 236,689) & \textbf{[0.550; 1.0]} & 4 & 8,769\\
\midrule
Head & \cite{zhang2013robust} & (64,258; 56,419) & [0.0130; 1.0] & 3 & \textbf{169}\\
(Genus-0)& ours & \textbf{(62,782; 55,038)} & \textbf{[0.550; 1.0]} & 5 & 444\\
\midrule
Lion Recon & \cite{gao2019feature} & (134,140; 115,245) & [0.178; 0.999] & 4 & 18,755\\
(Genus-0)& ours & \textbf{(112,847; 95,251)} & \textbf{[0.570; 1.0]} & 4 & \textbf{2,046}\\
\midrule
Oil Pump & \cite{gao2019feature} & \textbf{(75,033; 63,012)} & [0.117; 0.999] & 4 & 14,301\\
(Genus-4)& ours & (233,702; 196,455) & \textbf{[0.540; 1.0]} & 5 & \textbf{9,742}\\
\midrule
Ramses & \cite{gao2019feature} & (54,634; 46,923) & [0.0161; 0.999] & 4 & 9,395\\
(Genus-0)& ours & \textbf{(44,790; 37,993)} & \textbf{[0.590; 1.0]} & 4 & \textbf{713}\\
\midrule
Red Circular Box & \cite{gao2019feature} & (409,011; 367,583) & [0.215; 0.999] & 4 & 71,626\\
(Genus-0)& ours &\textbf{(351,881; 313,866)} & \textbf{[0.560; 1.0]} & 4 & \textbf{7,547}\\
\midrule
Thai Statue & \cite{gao2019feature} & (70,561; 59,470) & [0.180; 0.999] & 4 & 26,075 \\
(Genus-3)& ours & \textbf{(64,764; 53,831)} & \textbf{[0.550; 1.0]} & 4 & \textbf{1,845}\\
\bottomrule
\end{tabular}    
\end{table}

\section{Conclusion and Future Work}
\label{sec:4}

In this paper, we have presented HybridOctree\_Hex, a software package for adaptive all-hex mesh generation. Our approach leverages an octree structure to efficiently detect key surface features with curvature and thickness and construct a strongly balanced octree. From the strongly balanced octree, we directly generate an all-hex dual mesh using pre-defined templates, bypassing the implementation of sophisticated hybrid octree construction steps in \cite{hu2013adaptive}. Then we clear the buffer zone of the all-hex mesh with a face-normal-based surface improvement technique to obtain a core mesh with the $\min\mathit{SJ}$ of $0.258$. After that, the buffer zone is meshed by connecting core mesh boundary points with their corresponding points on the input surface. In the quality improvement step, we fit the outmost points to the input surface by iteratively minimizing a comphrehensive energy function composed of geometry fitting, scaled Jacobian, and Jacobian terms. Smart Laplacian smoothing is performed every $1,000$ iterations to smooth the outmost two layers of the vertices and drag surface points that are stuck in the local minima. Our empirical evaluation demonstrates the robustness and efficiency of HybridOctree\_Hex, as it handles dozens of complex 3D models without requiring manual intervention or parameter adjustment. Among all the models we tested, our algorithm generates meshes in the shortest time compared to \cite{hu2013adaptive, gao2019feature}, with high adaptation and the $\min\mathit{SJ}$ of $> 0.5$ after quality improvement. The source code and comprehensive statistical and meshing results are made available to facilitate further research and development in the field; see \url{https://github.com/CMU-CBML/HybridOctree_Hex}.

While HybridOctree\_Hex has demonstrated promising results in terms of fast, robust, and high-quality all-hex mesh generation, there remain several avenues for future research. One such direction is the exploration of more advanced techniques for curvature and narrow region detection to further improve the identification of key surface features. In our current approach, we employ pre-assigned thresholds, set conservatively to ensure the correct topology of the resulting mesh. However, this can lead to unnecessarily dense elements. A possible solution lies in introducing machine learning, which has already shown potential in 2D quad mesh generation \cite{tong2023srlafm}. Neural networks can be levaraged to predict the octree level distribution on the surface, learning from both successful and failed adaptive octree configurations. Another area of focus is the energy function employed in the optimization process. In theory, the energy terms can be expanded to generate customized hex meshes. For example, we can substitute the Jacobian and scaled Jacobian terms with alternative mesh quality metrics, enabling control over mesh quality from diverse perspectives, such as anisotropic mesh generation. Furthermore, the geometry fitting term can be strengthened to preserve user-defined sharp features by fixing corner points and constraining edge points to move exclusively along feature edges.  HybridOctree\_Hex has already exhibited superior efficiency compared to previous methods \cite{hu2013adaptive, gao2019feature}. Nevertheless, there is scope for further improvement in computational cost through the utilization of parallel computing, paving the way for real-time applications.

\section{Acknowledgment}

H. Tong and Y. J. Zhang were supported in part by the NSF grant CMMI-1953323 and a Honda grant.

\vspace{-3mm}
\bibliographystyle{elsarticle-num}
\bibliography{refs}

\begin{thebibliography}{10}
\expandafter\ifx\csname url\endcsname\relax
  \def\url#1{\texttt{#1}}\fi
\expandafter\ifx\csname urlprefix\endcsname\relax\def\urlprefix{URL }\fi
\expandafter\ifx\csname href\endcsname\relax
  \def\href#1#2{#2} \def\path#1{#1}\fi

\bibitem{zhang2016geometric}
Y.~J. Zhang, {Geometric Modeling and Mesh Generation from Scanned Images}, CRC Press, Taylor \& Francis Group, 2016.

\bibitem{benzley1995comparison}
S.~E. Benzley, E.~Perry, K.~Merkley, B.~Clark, G.~Sjaardama, A comparison of all hexagonal and all tetrahedral finite element meshes for elastic and elasto-plastic analysis, in: 4th International Meshing Roundtable, 1995, pp. 179--191.

\bibitem{shepherd2006quality}
J.~F. Shepherd, C.~J. Tuttle, C.~Silva, Y.~Zhang, Quality improvement and feature capture in hexahedral meshes, Technical Report UUSCI-2006-029, The University of Utah (2006).

\bibitem{owen1998survey}
S.~J. Owen, A survey of unstructured mesh generation technology, International Meshing Roundtable 239~(267) (1998) 15.

\bibitem{zhang2013challenges}
Y.~Zhang, Challenges and advances in image-based geometric modeling and mesh generation, Image-Based Geometric Modeling and Mesh Generation (2013) 1--10.

\bibitem{schneiders2000algorithms}
R.~Schneiders, Algorithms for quadrilateral and hexahedral mesh generation, Proceedings of the VKI Lecture Series on Computational Fluid Dynamic (2000).

\bibitem{blacker2000meeting}
T.~Blacker, Meeting the challenge for automated conformal hexahedral meshing, in: 9th International Meshing Roundtable, Citeseer, 2000, pp. 11--20.

\bibitem{tautges2001generation}
T.~J. Tautges, The generation of hexahedral meshes for assembly geometry: survey and progress, International Journal for Numerical Methods in Engineering 50~(12) (2001) 2617--2642.

\bibitem{shepherd2008hexahedral}
J.~F. Shepherd, C.~R. Johnson, Hexahedral mesh generation constraints, Engineering with Computers 24~(3) (2008) 195--213.

\bibitem{shepherd2007quality}
J.~F. Shepherd, Y.~Zhang, C.~Tuttle, C.~Silva, Quality improvement and {B}oolean-like cutting operations in hexahedral meshes, The 10th ISGG Conference on Numerical Grid Generation (2007).

\bibitem{gregson2011all}
J.~Gregson, A.~Sheffer, E.~Zhang, All-hex mesh generation via volumetric polycube deformation, Computer Graphics Forum 30~(5) (2011) 1407--1416.

\bibitem{hu2016centroidal}
K.~Hu, Y.~J. Zhang, Centroidal {V}oronoi tessellation based polycube construction for adaptive all-hexahedral mesh generation, Computer Methods in Applied Mechanics and Engineering 305 (2016) 405--421.

\bibitem{hu2017surface}
K.~Hu, Y.~J. Zhang, T.~Liao, Surface segmentation for polycube construction based on generalized centroidal {V}oronoi tessellation, Computer Methods in Applied Mechanics and Engineering 316 (2017) 280--296.

\bibitem{yu2022hexgen}
Y.~Yu, X.~Wei, A.~Li, J.~G. Liu, J.~He, Y.~J. Zhang, Hex{G}en and {H}ex2{S}pline: polycube-based hexahedral mesh generation and spline modeling for isogeometric analysis applications in {LS-DYNA}, in: Geometric Challenges in Isogeometric Analysis, Springer, 2022, pp. 333--363.

\bibitem{nieser2011cubecover}
M.~Nieser, U.~Reitebuch, K.~Polthier, Cubecover--parameterization of 3{D} volumes, Computer Graphics Forum 30~(5) (2011) 1397--1406.

\bibitem{li2012all}
Y.~Li, Y.~Liu, W.~Xu, W.~Wang, B.~Guo, All-hex meshing using singularity-restricted field, ACM Transactions on Graphics 31~(6) (2012) 1--11.

\bibitem{liu2018singularity}
H.~Liu, P.~Zhang, E.~Chien, J.~Solomon, D.~Bommes, Singularity-constrained octahedral fields for hexahedral meshing, ACM Transactions on Graphics 37~(4) (2018) 93--1.

\bibitem{zhang20053d}
Y.~Zhang, C.~Bajaj, B.-S. Sohn, 3{D} finite element meshing from imaging data, Computer Methods in Applied Mechanics and Engineering 194~(48-49) (2005) 5083--5106.

\bibitem{zhang2006adaptive}
Y.~Zhang, C.~Bajaj, Adaptive and quality quadrilateral/hexahedral meshing from volumetric data, Computer Methods in Applied Mechanics and Engineering 195~(9-12) (2006) 942--960.

\bibitem{zhang2006quality}
Y.~Zhang, G.~Xu, C.~Bajaj, Quality meshing of implicit solvation models of biomolecular structures, Computer Aided Geometric Design 23~(6) (2006) 510--530.

\bibitem{zhang2008automatic}
Y.~Zhang, T.~J. Hughes, C.~L. Bajaj, Automatic 3{D} meshing for a domain with multiple materials, in: 16th International Meshing Roundtable, 2008, pp. 367--386.

\bibitem{zhang2010automatic}
Y.~Zhang, T.~J. Hughes, C.~L. Bajaj, An automatic 3{D} mesh generation method for domains with multiple materials, Computer Methods in Applied Mechanics and Engineering 199~(5-8) (2010) 405--415.

\bibitem{qian2012automatic}
J.~Qian, Y.~Zhang, Automatic unstructured all-hexahedral mesh generation from {B}-{R}eps for non-manifold {CAD} assemblies, Engineering with Computers 28 (2012) 345--359.

\bibitem{marechal2009advances}
L.~Mar{\'e}chal, Advances in octree-based all-hexahedral mesh generation: handling sharp features, in: 18th International Meshing Roundtable, 2009, pp. 65--84.

\bibitem{hu2013adaptive}
K.~Hu, J.~Qian, Y.~Zhang, Adaptive all-hexahedral mesh generation based on a hybrid octree and bubble packing. 22nd {I}nternational {M}eshing {R}oundtable (2013).

\bibitem{gao2019feature}
X.~Gao, H.~Shen, D.~Panozzo, Feature preserving octree-based hexahedral meshing, Computer Graphics Forum 38~(5) (2019) 135--149.

\bibitem{livesu2021optimal}
M.~Livesu, L.~Pitzalis, G.~Cherchi, Optimal dual schemes for adaptive grid based hexmeshing, ACM Transactions on Graphics 41~(2) (2021) 1--14.

\bibitem{pitzalis2021generalized}
L.~Pitzalis, M.~Livesu, G.~Cherchi, E.~Gobbetti, R.~Scateni, Generalized adaptive refinement for grid-based hexahedral meshing, ACM Transactions on Graphics 40~(6) (2021) 1--13.

\bibitem{canann1998approach}
S.~A. Canann, J.~R. Tristano, M.~L. Staten, et~al., An approach to combined {L}aplacian and optimization-based smoothing for triangular, quadrilateral, and quad-dominant meshes, International Meshing Rountable 1 (1998) 479--94.

\bibitem{zhang2009surface}
Y.~Zhang, C.~Bajaj, G.~Xu, Surface smoothing and quality improvement of quadrilateral/hexahedral meshes with geometric flow, Communications in Numerical Methods in Engineering 25~(1) (2009) 1--18.

\bibitem{freitag1997combining}
L.~A. Freitag, On combining {L}aplacian and optimization-based mesh smoothing techniques, Technical Report, Argonne National Lab (1997).

\bibitem{qian2010quality}
J.~Qian, Y.~Zhang, W.~Wang, A.~C. Lewis, M.~S. Qidwai, A.~B. Geltmacher, Quality improvement of non-manifold hexahedral meshes for critical feature determination of microstructure materials, International Journal for Numerical Methods in Engineering 82~(11) (2010) 1406--1423.

\bibitem{lin2015quality}
H.~Lin, S.~Jin, H.~Liao, Q.~Jian, Quality guaranteed all-hex mesh generation by a constrained volume iterative fitting algorithm, Computer-Aided Design 67 (2015) 107--117.

\bibitem{meyer2003discrete}
M.~Meyer, M.~Desbrun, P.~Schr{\"o}der, A.~H. Barr, Discrete differential-geometry operators for triangulated 2-manifolds, Visualization and Mathematics III (2003) 35--57.

\bibitem{guigue2003fast}
P.~Guigue, O.~Devillers, Fast and robust triangle-triangle overlap test using orientation predicates, Journal of Graphics Tools 8~(1) (2003) 25--32.

\bibitem{zhang2013robust}
Y.~Zhang, X.~Liang, G.~Xu, A robust 2-refinement algorithm in octree or rhombic dodecahedral tree based all-hexahedral mesh generation, Computer Methods in Applied Mechanics and Engineering 256 (2013) 88--100.

\bibitem{paragios2002matching}
N.~Paragios, M.~Rousson, V.~Ramesh, Matching distance functions: a shape-to-area variational approach for global-to-local registration, in: 7th European Conference on Computer Vision, 2002, pp. 775--789.

\bibitem{rousson2002shape}
M.~Rousson, N.~Paragios, Shape priors for level set representations, in: 7th European Conference on Computer Vision, 2002, pp. 78--92.

\bibitem{tong2023srlafm}
H.~Tong, K.~Qian, E.~Halilaj, Y.~J. Zhang, S{RL}-{A}ssisted {AFM}: Generating planar unstructured quadrilateral meshes with supervised and reinforcement learning-assisted advancing front method, The Special Issue of ICCS AIHPC4AS in Journal of Computational Science 72 (2023) 102109.

\end{thebibliography}
\end{document}